\documentstyle[psfig,epsfig]{mn}

\newcommand{\be}{\begin{equation}}
\newcommand{\ee}{\end{equation}}
\newcommand{\ba}{\begin{eqnarray}}
\newcommand{\ea}{\end{eqnarray}}

\newcommand{\nn}{\nonumber \\}

\newcommand{\n}{\mbox{\boldmath $n$}}

\newcommand{\mnras}{MNRAS}
\newcommand{\prd}{Phys. Rev. D}

\newcommand{\apj}{Astro. Phys. Journal}
\newcommand{\aap}{AAP}

\def\gs{\mathrel{\raise1.16pt\hbox{$>$}\kern-7.0pt %         
\lower3.06pt\hbox{{$\scriptstyle \sim$}}}}         %         
\def\ls{\mathrel{\raise1.16pt\hbox{$<$}\kern-7.0pt %        
\lower3.06pt\hbox{{$\scriptstyle \sim$}}}}         %   

\title[RCSLenS: Cosmic Distances from Weak Lensing]{RCSLenS: Cosmic Distances from Weak Lensing}

\author[Kitching, et al.]
       {T. D. Kitching$^1$\thanks{t.kitching@ucl.ac.uk}, M. Viola$^2$, 
         H. Hildebrandt$^3$, A. Choi$^4$, T. Erben$^5$, D. G. Gilbank$^6$, 
         \newauthor
         C. Heymans$^4$, L. Miller$^7$, R. Nakajima$^8$, E. van Uitert$^9$ \\
        $^1$Mullard Space Science Laboratory, University College London, Holmbury St Mary, Dorking, Surrey RH5 6NT, UK\\
        $^2$Leiden Observatory, Leiden University, Niels Bohrweg 2, 2333 CA Leiden, The Netherlands\\
        $^3$Argelander Institute for Astronomy, University of Bonn, Auf dem Hugel 71, 53121, Bonn, Germany\\ 
        $^4$The Scottish Universities Physics Alliance, Institute for Astronomy, University of Edinburgh, Blackford Hill, Edinburgh EH9 3HJ, UK \\
        $^5$Argelander Institute for Astronomy, University of Bonn, Auf dem H (ugel 71, 53121 Bonn, Germany \\
        $^6$South African Astronomical Observatory, PO Box 9, Observatory, 7935, South Africa \\
        $^7$Department of Physics, University of Oxford, Denys Wilkinson Building, Keble Road, Oxford OX1 3RH, UK \\
        $^8$Argelander Institute for Astronomy, University of Bonn, Auf dem Hugel 71, 53121, Bonn, Germany\\ 
        $^9$Department of Physics and Astronomy, University College London, Gower Street, London, WC1E 6BT, UK}
       
\date{}

\pagerange{\pageref{firstpage}--\pageref{lastpage}}

\pubyear{2015}

\begin{document}

\maketitle

\label{firstpage}

\begin{abstract}
In this paper we present results of applying the shear-ratio method to the RCSLenS data. The 
method takes the ratio of the mean of the weak lensing tangential shear signal about galaxy clusters, 
averaged over all clusters of the same redshift, in 
multiple background redshift bins. In taking a ratio the mass-dependency of the shear signal is cancelled-out leaving 
a statistic that is dependent on the geometric part of the lensing kernel only. We apply 
this method to $535$ clusters and measure a cosmology-independent 
distance-redshift relation to redshifts $z\gs 1$. In combination with Planck 
data the method lifts the degeneracies in the CMB measurements, resulting in cosmological parameter constraints of  
$\Omega_M=0.31 \pm 0.10$ and $w_0 = -1.02\pm 0.37$, for a flat wCDM cosmology.
\end{abstract}


\section{Introduction}
\label{Introduction}
Weak gravitational lensing (or `weak lensing') is the phenomenon whereby the images 
of distant galaxies are distorted by a small amount as a result of mass perturbations along the line of sight. 
The distortions are a change in the third flattening (or third eccentricity; known as `ellipticity') of the 
galaxy images, and size. The change in ellipticity of an object is called `shear'. 
Every matter perturbation along any line of sight will cause weak lensing distortions. The distortions caused 
by large-scale-structure are known as cosmic shear. In the case that there is an individual lensing mass, 
for example a single galaxy, galaxy group or cluster, then the distortions caused are an 
elliptical alignment in the direction perpendicular (tangential) 
to the angular separation between the centre of mass and the lensed objects projected position. 
The amplitude of the distortion in this case is a function of the mass distribution of the lensing object 
and the geometry of the source-lens-observer configuration. These two terms are in general separable, in the case that 
the extent of the lens along the line of sight is much smaller than source-lens distance. Therefore 
by taking the \emph{ratio} of the amplitude of the lensing distortion at two different redshifts behind the lens
the mass-dependent term cancels leaving only a measure of the geometry of setting. 

This approach to extracting 
the geometric part of the weak lensing signal is known as the `shear-ratio' method. It was first proposed 
in Jain \& Taylor (2003), and subsequently developed in Taylor et al. (2007) and Kitching et al. (2008). 
It has been applied to data twice in Kitching et al. (2007) and Taylor et al. (2012), but only on very small 
data sets as a proof of concept, it was also discussed in Medezinski et al. (2011). In this paper 
we apply this method to one of the largest combined weak lensing and cluster 
catalogues yet available, RCSLenS, and use this to measure the distance-redshift relation from weak lensing data alone. 

Purely geometric measurements of expansion history of the Universe, such as the supernovae magnitude-redshift 
relation and the Baryon Acoustic Oscillation (BAO) length-redshift relation, have played a critical part in 
building the standard cosmological model (e.g Samushia et al., 2014; Amanullah et al., 2010). 
They provide low-redshift complementary distance measures 
to the high-redshift Cosmic Microwave Background (CMB) allowing consistency in the cosmological model to be tested across the 
expansion history. Other probes such as cosmic shear, and measurements of the matter power spectrum from 
galaxy clustering, measure combinations of the geometry of the Universe and the growth of structure. In 
general dark energy models, and in particular for several modified gravity models, these two aspects may 
be affected in different ways (e.g. Shafieloo et al., 2013; Clifton et al., 2012). 
Therefore having both purely geometric methods, such as BAO, 
and geometry-plus-growth methods, such as galaxy clustering, results in powerful combinations in determining the 
nature of dark energy. The shear-ratio method provides a way for the geometric signal to be extracted from 
weak lensing data, which is complementary to cosmic shear. 

The shear-ratio method can be used to constrain cosmological parameters (Taylor et al., 2007), and also 
systematic effects in the weak lensing data (Kitching et al., 2008); in Heymans et al. (2012) and Kuijken et al. (2015) a 
similar method was used on galaxy scales 
(but one that did not take a ratio, and therefore was not a purely geometric method) 
to test photometric redshift systematic effects. In this paper we will use the lensing signal behind 
galaxy clusters to measure the distance-redshift relation, and infer cosmological parameters. 
We present systematic tests on the data and tests on simulated data to validate the method. 

The structure of this paper is as follows. In Section \ref{Method} we outline the methodology and data that 
is used. In Section \ref{Results} we present results, and cosmological parameter inference. In 
Section \ref{Conclusion} we present our conclusions. 

\section{Method}
\label{Method}
We use the approach outlined in Taylor et al., (2007) and Kitching et al. (2008). 
The raw data that we use is a 
catalogue of ellipticity and photometric redshift measurements from galaxies 
and a set of cluster positions and redshifts. 
For each cluster (`lens') we can identify those galaxies with a peak posterior photometric redshift greater than the cluster 
(`background galaxies'). For each galaxy we compute its projected tangential (perpendicular) $e^T$,  
and cross (45 degree) component $e^X$, ellipticities with respect to the angular separation 
between the cluster centre $(\theta^C_1$, $\theta^C_2)$ and the background galaxy $(\theta_{1,g}$, $\theta_{2,g})$
\ba
\label{ets}
e^T_g(\theta, z_i)&=&-(e_{1,g}\cos[2\phi_g]+e_{2,g}\sin[2\phi_g])\nn
e^X_g(\theta, z_i)&=& (e_{1,g}\sin[2\phi_g]-e_{2,g}\cos[2\phi_g]),
\ea
where $\phi_g=\tan^{-1}([\theta^C_1-\theta_{1,g}]/[\theta^C_2-\theta_{2,g}])$ is the angular position around the cluster, 
and $e_{1,g}$ and $e_{2,g}$ are the measured ellipticities in Cartesian 
coordinates on the sky, $g$ labels a single galaxy with 
local radial angular coordinate $\theta=\sqrt{(\theta^C-\theta_{1,g})^2+(\theta^C_2-\theta_{2,g})^2}$ and redshift $z$. 
Throughout we label source redshift bins using $i$ and $j$, and lens redshift bins using $L$. 

For a given lens we can divide the background population into a series of redshift bins, selected using the {\tt Z\_B} 
parameter of BPZ (Benitez, 2000; applied in Hildebrandt et al., 2015), which 
is the peak of the posterior probability distribution, 
to create a set of populations labelled by 
the lens redshift $z_L$, source background redshift $z_i$ and angular position: $\{e^T_g(\theta, z_i|z_L)\}$. We assume 
that, over a sufficiently large population of galaxies, the ellipticity is an unbiased estimate of the shear 
$\gamma^T\simeq \langle e^T\rangle$ (we will later account for the finite number of galaxies with a shot noise term 
in a covariance matrix of the measurement). 

\subsection{Theory}
\label{Theory}
We assume that the weak lensing shear signal, for a given lens, is a function of the lens mass distribution $M_L$, radial 
projected coordinate $\theta$, and a geometric part of the signal $G(z_L, z_i)$. We will further assume that these 
are related in the following general manner 
\be 
\label{tt}
\gamma^T(z_i,z_L|M_L,\theta)=f(M_L,\theta)G(z_i,z_L)
\ee
i.e. that the radial and mass-dependence $f$ is separable from the geometric part $G$. 
This is the case for both the Singular Isothermal Sphere (SIS) radial profile and the NFW profile (Navarro et al., 1996)
for a single lens. 
For the SIS case the functional form of the shear signal, for a lens at redshift $z_L$, with mass $M_L$, 
computed using background sources in a redshift bin $\Delta z_i$ is 
\ba 
\label{sis}
\gamma^T(z_i,z_L|M_L,\theta)
=f(M_L)\left(\frac{1}{\theta}\right)\frac{\int_{z\in \Delta z_i}{\rm d}z G(z,z_L)n(z|z_i)}{\int_{z\in \Delta z_i}{\rm d}z n(z|z_i)},
\ea
where the integral limit $z \in \Delta z_i$ means the integral over all redshifts in bin $i$, 
with weighted mean redshift $z_i$, 
with limits $z-\Delta z_i$ and $z+\Delta z_i$. 
All sums are within the bins, and the ``leakage'' between bins is accounted for in the integrals over 
the posterior redshift probability distributions. 
$n(z|z_i)=\sum_{g\in z_i}w_g p_g(z|z_g)/\sum_g w_g$ is the normalised sum of the posterior redshifts $p_g(z|z_g)$ 
for each galaxy, 
at a redshift $z_i$ i.e. 
the redshift distribution; where $w_g$ are weights for each galaxy provided in the catalogue (see 
Miller et al., 2013) that are related to the measured shape of the galaxies and are approximately 
an inverse variance. 

The geometric kernel $G$ is
\be
G(z_i,z_L)\equiv \frac{D(z_i)-D(z_L)}{D(z_i)D(z_L)}
\ee
where we have assumed a flat geometry, and $D(z)$ are comoving distances.
The comoving distance is given, for an observer at $z=0$, by 
\be 
\label{Dz}
D(z)=\int_0^z {\rm d}z' \frac{c}{H(z')}.
\ee
where $c$ is the speed of light in a vacuum. The 
Hubble parameter $H(z)$, for a flat geometry and a constant dark energy equation of state $w_0=p_{\rm de}/(c^2\rho_{\rm de})$ (where $p_{\rm de}$ and 
$\rho_{\rm de}$ are the pressure and density of the dark energy fluid respectively), is given by 
\be
H(z)=H_0[\Omega_{\rm M}(1+z)^3+(1-\Omega_{\rm M})(1+z)^{-3(1+w_0)}]^{1/2}, 
\ee
where $\Omega_{\rm M}$ is the current dimensionless matter overdensity, $H_0$ is the 
current value of the Hubble parameter. 

If we now bin the lenses in a redshift bin $\Delta z_L$, assumming that the mean masses of the lenses do not evolve inside 
$\Delta z_L$ then equation (\ref{sis}) is changed to 
\ba
\label{A1}
\langle\gamma^T(z_i,z_L|M_L,\theta)\rangle=\langle f(M_L)\rangle_{\Delta z_L}\left(\frac{1}{\theta}\right)\nn
\frac{\int_{z'_L\in \Delta z_L}{\rm d}z'_L q_L(z'_L)\int_{z\in \Delta z_i}{\rm d}z G(z,z'_L)n(z|z_i)}{\int_{z'_L\in \Delta z_L}{\rm d}z'_L q_L(z'_L)\int_{z\in \Delta z_i}{\rm d}zn(z|z_i)},
\ea
where $q_L(z)$ is the redshift distribution of the lenses in bin $\Delta z_L$, which is the 
observed number density of the lenses $n_L(z)$ smoothed by the photometric redshift error distribution $p_z(z'|z)$ 
of the lenses: $q_L(z)=\int {\rm d}z' n_L(z') p_L(z'|z)$.  
In this paper we use the photometric redshift error distribution of the galaxy cluster sample given in van Uitert et al. (2015) 
who quote a redshift error of $\sigma_{z_L}(z)=0.03$ 
assuming a Gaussian probability density function. 
$\langle f(M_L)\rangle_{\Delta z_L}=\sum_{L\in \Delta z_L} f(M_L)W(M)/\sum_L W(M)$ is the mean mass dependency in the 
bin $\Delta z_L$ with some arbitrary mass-dependent 
weight function $W(M)$ that we discuss further in Section \ref{Data}. We combine the amplitude into a single 
function $\langle A(z_i,z_L)\rangle_{\rm theory}$ where 
\be 
\label{ttt}
\langle\gamma^T(z_i,z_L|M,\theta)\rangle=\frac{\langle A(z_i,z_L)\rangle_{\rm theory} \langle f(M_L)\rangle_{\Delta z_L}}{\theta} 
\ee
and $\langle A(z_i,z_L)\rangle_{\rm theory}$ can be equated through comparison with equation (\ref{A1}).
This general expression is also true for an NFW profile except that the angular dependence is more complicated. Indeed
this separation of mass, geometric and angular parts 
is generally true for the majority of plausible symmetric lens configurations (Bartelmann \& Schneider, 2001).

In deriving this expression we assume that the sum over lenses $L$ is performed on thin slices $S$ 
in redshift within a bin $B$,  
$\sum_{S\in B} w_S(z_S) \sum_{L\in S} f(M_L)G(z_i,z_L)$ (where $G$ is the geometric kernel in equation \ref{tt}), so that in 
the limit that the slice is thin all lenses are at the same redshift and the geometric term drops out of the sum over 
lenses $\sum_{S\in B} w_S(z_S) G(z_i, z_S) \sum_{L\in S} f(M_L)$; by taking the continuous limit of this expression and weighting 
by the lens number density and probability distributions, we derive equation (\ref{ttt}). 
This results in the sums over lenses on each redshift slice becoming a mean 
computed for each lens redshift slice within the bin $\langle f(M_L)\rangle_L(z)=\sum_{L\in z} f(M_L)W(M)/\sum_L W(M)$. 
In order to proceed with taking ratios such that the mean mass distribution of the lenses can be cancelled-out we 
therefore need to assume that the mean of the lens mass as a function of redshift, within each lens bin, is constant. 

The constancy of the mean mass within narrow redshift slices is well justified through measurements of mass and richness as 
as function of redshift (see e.g. van Uitert et al., 2015) which only vary slowly, however 
we test the robustness of this assumption by varying the lens bin width in our analysis. 
By taking thinner bins in lens redshift the assumption that the mean mass dependency is constant across the bin will be more 
accurate and the number of data points will increase, however the noise per bin will get larger and the contamination between 
bins due to the photometric scatter will also get larger. 
In Appendix A we show the final cosmological parameter constraints 
presented in Section \ref{Results} as a function of lens bin width. In the main analysis we use a bin width of 
$\Delta z_L=0.18$, which has the fewest number of bins in the redshift range that we consider and should be most robust to 
the lens photometric redshift probability distribution. 

\subsection{Ratios}
To extract the geometric information from each lens we fit the function $A(z_i,z_L)/\theta$ to each source redshift bin,
using a $\chi^2$-minimisation on the set of galaxies $\{e^T_g(\theta, z_i|z_L)\}$, as we describe in more detail in 
Section \ref{Data}.  
This provides a set of
amplitudes $\{A(z_i,z_L)\}$ which are estimates of the numerator in equation 
(\ref{ttt}) for each lens. 
From the data we can compute the mean over all lenses for each lens-source redshift bin pair,
of the SIS amplitudes 
\be 
\langle A(z_i,z_L)\rangle_{\rm Data}=\frac{\sum_L A(z_i,z_L) W(M)}{\sum_L W(M)},
\ee 
which has the theoretical prediction given by equation (\ref{ttt}).

For all lenses within the same redshift bin we can now define a ratio 
\be
R^L_{ij}=\frac{\langle A(z_i,z_L)\rangle_{\rm Data}}{\langle A(z_j,z_L)\rangle_{\rm Data}},
\ee
where $z_j>z_i$, which has the expected theoretical value given by equation (\ref{ttt}), and 
\be 
\label{theory}
T^L_{ij}=\frac{\langle A(z_i,z_L)\rangle_{\rm theory}}{\langle A(z_j,z_L)\rangle_{\rm theory}},
\ee
where the mass dependency is cancelled in taking the ratio, and we are left with a statistic that depends only 
on ratios of comoving distances i.e. geometry. Note that we have taken the ratio of the mean of two quantities, not 
the mean of a ratio, and that therefore this statistic is not prone to divergences such as those discussed in e.g. 
Viola, Kitching, Joachimi (2014). 

\subsection{Covariance} 
To construct a likelihood function for the ratios we need to specify their covariance. This is composed 
of three terms as described in Taylor et al. (2007) and Kitching et al. (2008). 
The first is a shot noise term, whose fractional covariance is given by  
\ba
&&\frac{\langle R^L_{ij}R^L_{mn}\rangle_{\rm SN}}{R^L_{ij}R^L_{mn}}=
\left(\frac{\sigma^2_e}{2[\gamma^T(z_i,z_L)]^2}\right)(\delta^K_{im}-\delta^K_{in})+\nn
&&\left(\frac{\sigma^2_e}{2[\gamma^T(z_j,z_L)]^2}\right)(\delta^K_{jn}-\delta^K_{jm}),
\ea
where $\sigma^2_e$ is the mean per-component variance of the background galaxy ellipticities which is estimated from 
the data, $\gamma^T(z_i,z_L)$ is an estimate of the shear at redshift $z_i$ (equation \ref{sis}), and $\delta^K$ are Kronecker 
delta functions, and $ij$ and $mn$ are background redshift bin pairs. The second term is due to matter perturbations 
along the line of sight between the background redshift bins and the lens, the `cosmic shear' noise (Hoekstra, 2001), 
\ba
\frac{\langle R^L_{ij}R^L_{mn}\rangle_{\rm CS}}{R^L_{ij}R^L_{mn}}&=&\nn
\frac{C^1_{im}}{\gamma^T(z_i,z_L)\gamma^T(z_m,z_L)}+
\frac{C^1_{jn}}{\gamma^T(z_j,z_L)\gamma^T(z_n,z_L)}&-&\nn
\frac{C^1_{in}}{\gamma^T(z_i,z_L)\gamma^T(z_n,z_L)}-
\frac{C^1_{jm}}{\gamma^T(z_j,z_L)\gamma^T(z_m,z_L)}&+&\nn
\frac{C^2_{i,{\rm min}(j,n)}}{[\gamma^T(z_j,z_L)]^2}\delta^K_{im}+
\frac{C^2_{{\rm max}(i,m),j}}{[\gamma^T(z_j,z_L)]^2}\delta^K_{jn}.
\ea
The first four terms are due to the
correlated distortions induced on both background galaxy redshift bins 
by matter lying in front of the nearest source plane.
The last two terms arise from matter lying between the background
source planes and should be regarded as an extra noise
term on the ellipticities of the furthest background source
galaxies. The $C^1$ and $C^2$ functions are given by 
\be 
C^{\alpha}_{ij}=\int_0^{\infty}\frac{\ell{\rm d}\ell}{\pi}C^{\alpha}_{ij}(\ell)
\left\{\frac{2[1-J_0(\ell\theta_{\rm max})]}{\ell^2\theta^2_{\rm max}}-\frac{J_1(\ell\theta_{\rm max})}{\ell\theta_{\rm max}}\right\}^2
\ee
where $\alpha=[1,2]$, $\theta_{\rm max}$ is the angular size of the aperture in which the 
ellipticities are used about the lens, and $J_i$ are Bessel functions of the first kind. 
The power spectrum functions, assuming the Limber approximation, are given by 
\ba 
C^1_{ij}(\ell)&=&{\mathcal A}\int_0^{D(z_i)}{\rm d}D P(\ell/D[z]; D[z])\nn
&&{\mathcal W}(D,D(z_i)){\mathcal W}(D,D(z_j))\nn
C^2_{ij}(\ell)&=&{\mathcal A}\int_{D(z_i)}^{D(z_j)}{\rm d}D P(\ell/D[z]; D[z])\nn
&&{\mathcal W}^2(D,D(z_j)),
\ea
where $P(\ell/D; D[z])$ is the power spectrum of matter overdensity perturbations at a 
radial scale $k=\ell/D$ and redshift $z$, and $D(z)$ is given by equation (\ref{Dz}). 
The pre-factor is ${\mathcal A}=9\Omega^2_{\rm M}H^4_0/4c^4$. 
The weight function is 
\be 
{\mathcal W}(D,D[z_j]|D(z_L))=\int_{z_L}^{\infty} {\rm d}z \frac{n(z|z_j)}{a(z)}\frac{(D-D(z_L))D(z_j)}{(D(z_j)-D(z_L))D}
\ee
where $a(z)$ is the dimensionless scale factor at redshift $z$. 

The third term of the covariance is due to intrinsic alignment effects, which is given by equation (21) in Kitching et al. 
(2008), however its impact is expected to be negligible so we do not reproduce this here. 
Therefore the total covariance is given by 
\be 
C^L_{\mu\nu}=\langle R^L_{\mu}R^L_{\nu}\rangle_{\rm SN}+\langle R^L_{\mu}R^L_{\nu}\rangle_{\rm CS}
\ee
for each lens redshift $z_L$, and background redshift $\mu=ij$ and $\nu=mn$ pair combinations.

\subsection{Likelihood}
We can now construct a log-likelihood function for the ratio, assuming Gaussian distributed data 
\be 
\label{like}
-2\ln{\mathcal L}(\Theta)=\sum_L\sum_{\mu\nu}
(R^L_{\mu}-T^L_{\mu}(\Theta))
[C^L_{\mu\nu}]^{-1}
(R^L_{\nu}-T^L_{\nu}(\Theta))
\ee
where we neglect a constant additive factor. $\Theta$ are parameters of interest (i.e. cosmological 
parameters, or otherwise) and the sums run over all lens redshifts labelled by $L$, and background redshift bin pairs $\mu$ and 
$\nu$. 

In practice some redshift bin pairs are degenerate, for example the pairs $(i,i+1)$, $(i+1,i+2)$ and 
$(i,i+2)$ where the third pair could be constructed from a combination of the first two. To avoid this 
we take a complete set of redshift bin pairs to be only those $R^L_{ij}$ where $j=i+1$; or $\mu=(i,i+1)$. 
This also has the advantage that the covariance matrix is a band-diagonal matrix 
and so numerically stable to invert. 

In general a further weight function can be applied to the data and theory, $R^L_{ij}H(z_i,z_j,z_L)$, $T^L_{ij}H(z_i,z_j,z_L)$, 
where $H$ is some arbitrary function of the lens and background galaxy redshifts. A particularly convenient weight function is 
the geometric distance relation for a fiducial cosmology $H(z_i,z_j,z_L)=[D(z_j)-D(z_L)]/D(z_j)$, which in the case that the 
real cosmology were equal to the fiducial would result in $T^L_{ij}$ being a function of the redshift $z_i$ only.   
We test the likelihood code using mock ratio data as described in Appendix B. 

\subsection{Data}
\label{Data}
The data we use is the RCSLenS (Hildebrandt et al., in prep) data which is a re-analysis of the RCS2 data 
(Gilbank et al. 2011); observed with the CFHT (Canada France Hawaii Telescope). 
RCS2 was $785$ pointings. $765$ out of those have $r$-band data, and $761$ out of those have been successfully 
processed for RCSLenS. This corresponds to an effective area after masking of $571.8$ square degrees if only 
the $r$-band masks are used. $513$ out of the $761$ fields have $g$, $r$, $i$, $z$ coverage, 
this corresponds to an effective area of $383.5$ square degrees, now using masks from all filters, which is what we 
use in this paper. The RCS2 survey has been used several times to study galaxy cluster properties for example 
Sharon et al. (2015), Hoag et al. (2015), Anguita et al. (2012), Wuyts et al. (2010). 
In addition there are several papers based on the RCS2 data including galaxy-galaxy 
lensing (van Uitert et al., 2011; Cacciato et al., 2014) and galaxy clustering 
(Blake et al., 2015). An overview of the survey, data quality, photometric accuracy, etc. is given in Gilbank et al. (2011) 
and additional details for the specific applications may be found in the other papers listed.
The cluster catalogue, based on the Gilbank et al. 2011 photometric catalogues, is outlined in van Uitert et al. (2015). 
Briefly, this applied a simplified version of a red-sequence cluster-finder, 
largely as described in Lu, Gilbank, Balogh, Bognat (2009).
The data was reduced using the {\tt THELI} pipeline (Erben et al., 2012) 
with the shape measurement ellipticities and weights described in Miller et al. (2012), 
created using {\tt lensfit} (Miller et al., 2007; Kitching et al., 2008).

We use all the available RCSLenS data and do not make a pass/fail selection based on 
cosmic shear statistics as described in Hildebrandt et al. (in prep) adapted from Heymans et al. (2012), 
as these refer to shear-shear statistics, 
and they aim at detecting residual coherent PSF ellipticity in the shear catalogue, 
while we are employing here a position-shear statistic which is much less sensitive to residual PSF ellipticity 
in the shear measurements, as it is done by azimuthally averaging the shear measurements around each lens. For this 
analysis the cross-component ellipticity analysis is the relevant systematic test. The pass/fail test also 
The catalogoues were created in four different versions that employed a simple blinding scheme, such that four 
versions of this paper were created with the true catalogue unknown, then an unblinding was performed before submission. 

In equation (\ref{A1}) we use a weight function $W(M)$ that is the signal-to-noise of the cluster detection; 
although results are not sensitive to this choice, we reran the analysis using no weighting and the 
cosmological parameter results were unchanged. We use only those clusters 
and groups with a signal-to-noise of $10$ or more, resulting in a final catalogue of $535$ clusters which have 
background galaxies available for a lensing analysis in all background redshift bins.  

Around each cluster we take a maximum angular extent corresponding to a comoving transverse separation of 
$1h^{-1}$Mpc (approximately corresponding to $R_{200}$, see van Uitert et al., 2015), 
computed using a Planck best-fit cosmology (Planck Collaboration, 2014). We use a minimum angle 
corresponding to $0.01h^{-1}$Mpc to minimize the effect of a possible miscentre of the cluster position with regard to 
the true centre of the dark matter cluster halo, as well as contamination from central bright galaxy light in 
measurement of shapes of surrounding galaxies. We use $5$ background galaxy redshift bins, approximately 
corresponding to the maximum redshift divided by the mean photometric redshift error $\sim 1.5/0.2$, 
that are centred on $z_i=[0.39, 0.57, 0.75, 0.93, 1.11]$. 
For RCSLenS the photometric redshifts are known to have a large scatter for {\tt Z\_B} $< 0.4$, but this regime is avoided in our source 
selection except for the lowest redshift bin. 

\begin{figure}
    \includegraphics[angle=0,clip=,width=\columnwidth]{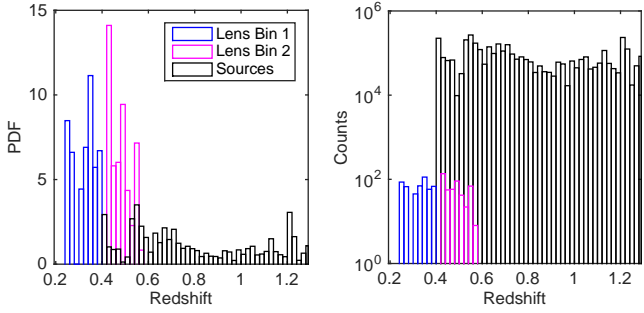}
 \caption{The distribution of the clusters in each of the bins used, and the 
   weak lensing sources in the RCSLenS data with a redshift greater than the minimum cluster redshift in the catalogues 
   $z>0.241$. The mean redshift of the lens bins are $0.33$ and $0.5$. The left panel shows the probability density 
   function of the distributions, the right panel shows the number counts (on a logarithmic y axis to account for 
   much larger number of source background galaxies than lenses).}
 \label{distributions}
\end{figure}

We use two cluster redshift bins 
of $0.24\leq z_L<0.40$, and $0.4\leq z_L <0.6$, where the minimum cluster redshift in the catalogue is $0.241$. 
Taking such large bins in lens-redshift decreases the signal in 
the measurement, but the larger number of background galaxies in each lens-source redshift 
bin combination also decreases the shot noise contribution. For the analysis we sum over lenses in each bin, 
and include the redshift error from van Uitert et al. (2015) as described in equation (\ref{ttt}). 
In Figure \ref{distributions} we show the probability 
density function of the distribution of clusters in each bin as well as the distribution of weak lensing sources 
(taking into account the RCSLenS shape measurement weight, see Hildebrandt et al., in prep). 

Using this data we then fit the function $A(z_i,z_L)/\theta$ to the set of tangential or cross 
component ellipticities in each bin, given in equation (\ref{emean}) 
to find the best fit set of amplitudes. This is done using 
a $\chi^2$-minimisation where for each source redshift bin for each lens we divide the source galaxies 
into $5$ angular bins $B$ corresponding to projected comoving separations of $B=[0.01, 0.02, 0.1, 0.2, 1.0]h^{-1}$Mpc 
(which are the centre of the radial bins). We explicitly assume here that the SIS is a good fit to the data. The fitting is 
done for each lens-background redshift bin pair 
separately, and then the mean amplitude over all lenses 
for each background redshift bin taken as described in Section \ref{Theory}. 
In each bin we compute the mean tangential and cross-component ellipticities 
\be 
\label{emean}
\langle e^T\rangle_B=
\frac{\sum_{g\in B}e^T_{g,{\rm corrected}}w_{\rm LF}}{\sum_{g\in B}w_{\rm LF}}
\frac{\sum_{g\in B}m^0_g w_{\rm LF}}{\sum_{g\in B}m^T_g w_{\rm LF}}
\ee
where the `corrected tangential ellipticity' is corrected for the additive systematic residuals $c_i$ 
(see Miller et al., 2013 
for an exposition of this, and Hildebrandt et al., in prep, for details of the RCSLenS derived values), the 
average value of the $c$ term over all lenses and background source redshift bins is $0.0016$. The additive 
residuals are simply subtracted from the observed ellipticities as follows
\be 
e^T_{g,{\rm corrected}}=-((e_{1,g}-c_{1,g})\cos[2\phi_g]+(e_{2,g}-c_{2,g})\sin[2\phi_g]).
\ee
We also correct for multiplicative systematic effects, as described in Miller et al. (2013). This is achieved 
by replacing the observed ellipticity with a $(1+m_{i,g})$ term in the statistic used, then applying a weight 
so that in the limit of no systematic ($m_{i,g}=0$) the result is unchanged. This is done in equation (\ref{emean}) 
where the multiplicative correction factors are 
\ba 
m^T_g&=&-((1+m_{1,g})\cos[2\phi_g]+(1+m_{2,g})\sin[2\phi_g])\nn
m^0_g&=&-(\cos[2\phi_g]+\sin[2\phi_g]).
\ea
A similar calculation is done for the cross-component part, see equation (\ref{ets}), and for 
the variance of the ellipticities we take the mean of the square of the $e^T_{g,{\rm corrected}}$ which is 
$\sigma^2_{e^T}=0.06$. Note that in our case $m_1\equiv m_2$ and that, because we have a high background number of 
galaxies that fairly sample $\phi=[0, 2\pi]$, the sums over the multiplicative bias reduce to approximately 
$\sum_{g\in B} w_{\rm LF}/\sum_{g\in B}(1+m) w_{\rm LF}$. The average value of this quantity is 
$1.06$ over the set of lenses and background source redshift bins, however we 
use the full expression to avoid any assumptions. 
We then take the mean over all clusters in each lens redshift bin, and then compute the set of ratios 
$R^L_{ij}$, where $j=i+1$. 

In taking a ratio any constant multiplicative bias also cancels out, so that the above 
correction is to account for any redshift-dependence of this bias. However because we only take ratios of neighbouring 
bins the method should also be relatively insensitive to these 
redshift dependent terms (unless the bias changes by a large amount 
over the redshift interval between bins, which is not expected). For example if we expand the multplicative bias 
to be $m(z)=m_0+m_1(1+z)$ then to linear order in $m_1$ the measured ratio becomes 
$[1+(m_1/m_0)(z_i-z_j)]R^L_{ij}$, 
if we have $m_1\ls 10^{-3}$, $m_0\ls 10^{-2}$ 
and $(z_i-z_j)\simeq 0.2$ then the correction is a $\ls 2\%$ change in the ratio. 
The ratio is also expected to be insensitive to changes in the additive bias as discussed in Kitching et al. (2008) where 
it was shown that uncertainties in the dark energy equation parameter scale approximately 
as $\delta w_0 = 0.4 (\delta c/0.01)$ where $\delta c$ is an unknown uncertainty in the additive bias.  
Therefore for changes of $\delta c \ls 10^{-3}$, which is an order of 
magnitude estimate expected for the RCSLenS shape measurement 
(Miller et al., 2013), the contribution to the error on cosmological parameters 
from misestimation of the additive bias are expected to be $\delta w_0\ls 0.04$. 

\begin{figure*}
    \includegraphics[angle=0,clip=,width=\columnwidth]{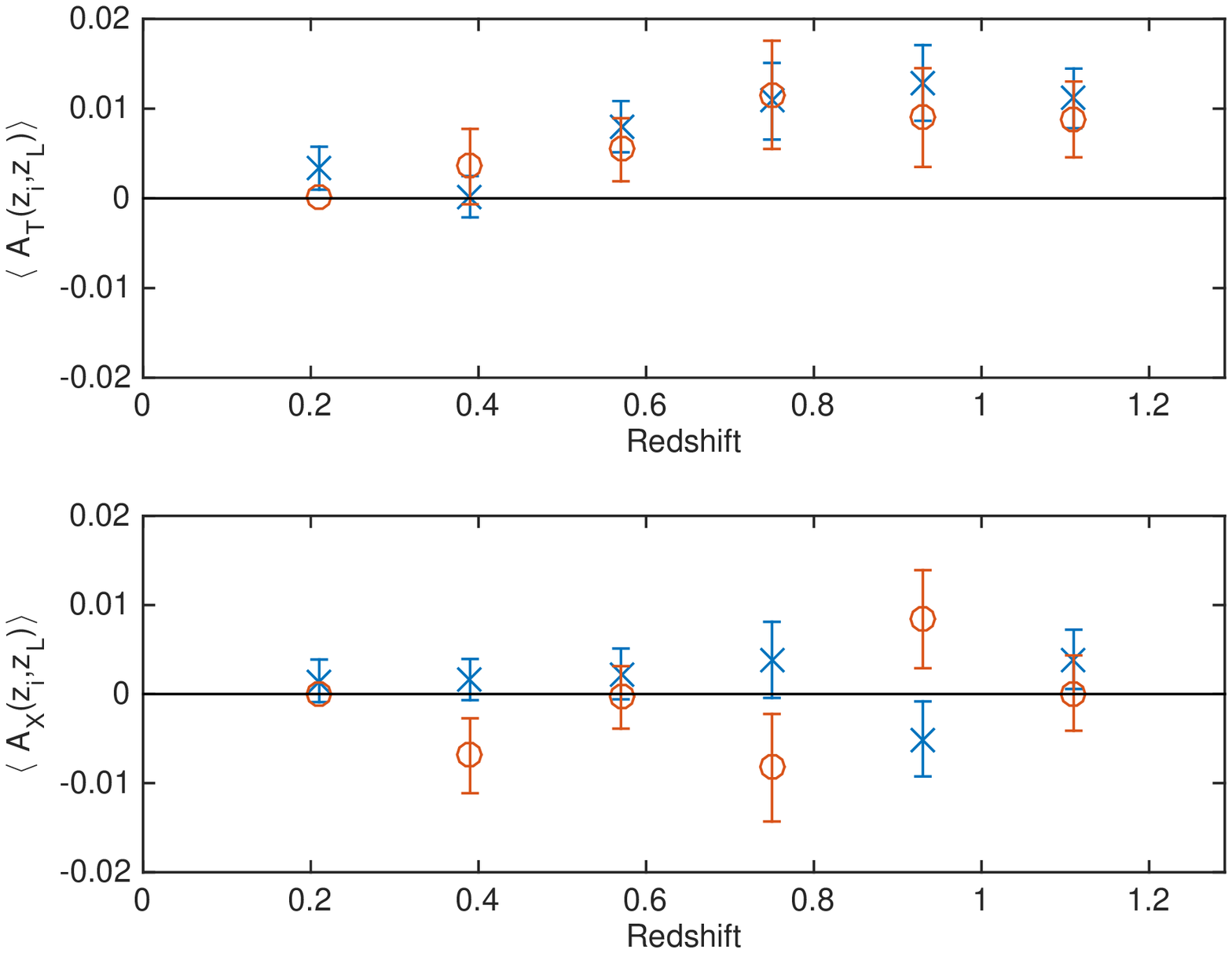}
    \includegraphics[angle=0,clip=,width=0.97\columnwidth]{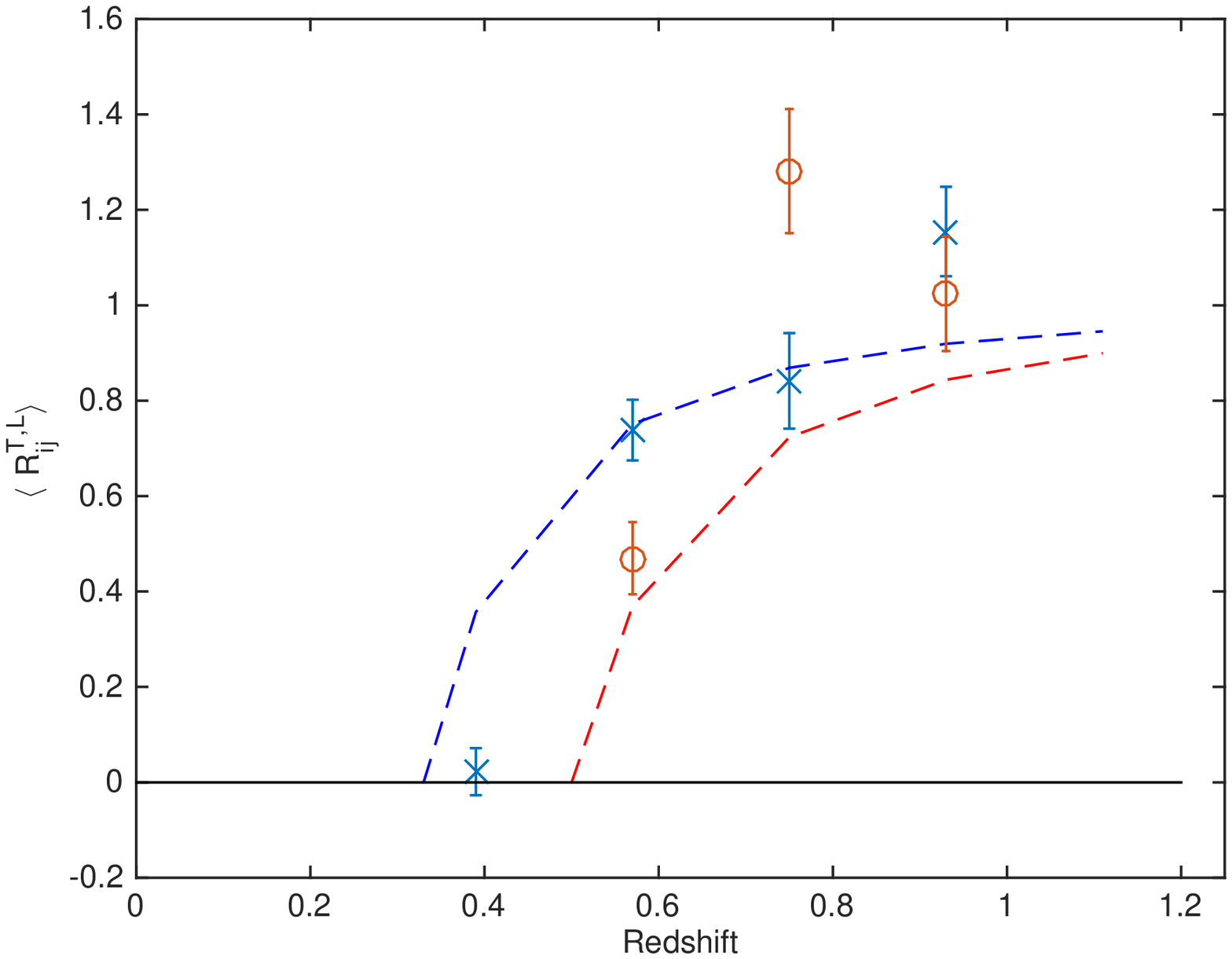}
 \caption{Left: The measured shear amplitudes fitting to the tangential 
(upper panel) and cross-component (lower panel) ellipticities averaged over all clusters in 
redshift bins $0.24\leq z_L<0.40$ (blue, marked `x'), and $0.40\leq z_L <0.60$ (red, marked `o'). 
Right: The ratio of the 
shear amplitudes $R_{ij}=\langle A_T(z_i,z_L)\rangle/\langle A_T(z_j,z_L)\rangle$ 
where $j=i+1$. We show the prediction for a Planck best fit cosmology (Planck Collaboration, 2014) 
for the two lens redshifts (equation \ref{theory}).}
 \label{ratios}
\end{figure*}

In Figure \ref{ratios} we show the measured tangential $A_T$, and cross component 
$A_X$, mean amplitudes. 
The cross component is consistent with zero, and is a systematic test of the measurement. 
We also show the measured ratios from these amplitudes. The signal $R^{T,L}_{ij}$ is significantly 
non-zero at $4.2$-sigma for the two lens redshift bins. 
We do not show the ratio of the $A_X$ amplitudes because the values are consistent with (very close to zero), 
so that the distribution of the ratio is divergent. 
In this case the $A_X$ are the relevant systematic quantity. 
This is also why a ratio is not plotted in the $z\leq z_L$ regime where the theoretical ratio signal is also divergent.

\section{Results}
\label{Results}
\begin{figure*}
\centering
  \includegraphics[angle=0,clip=,width=\columnwidth]{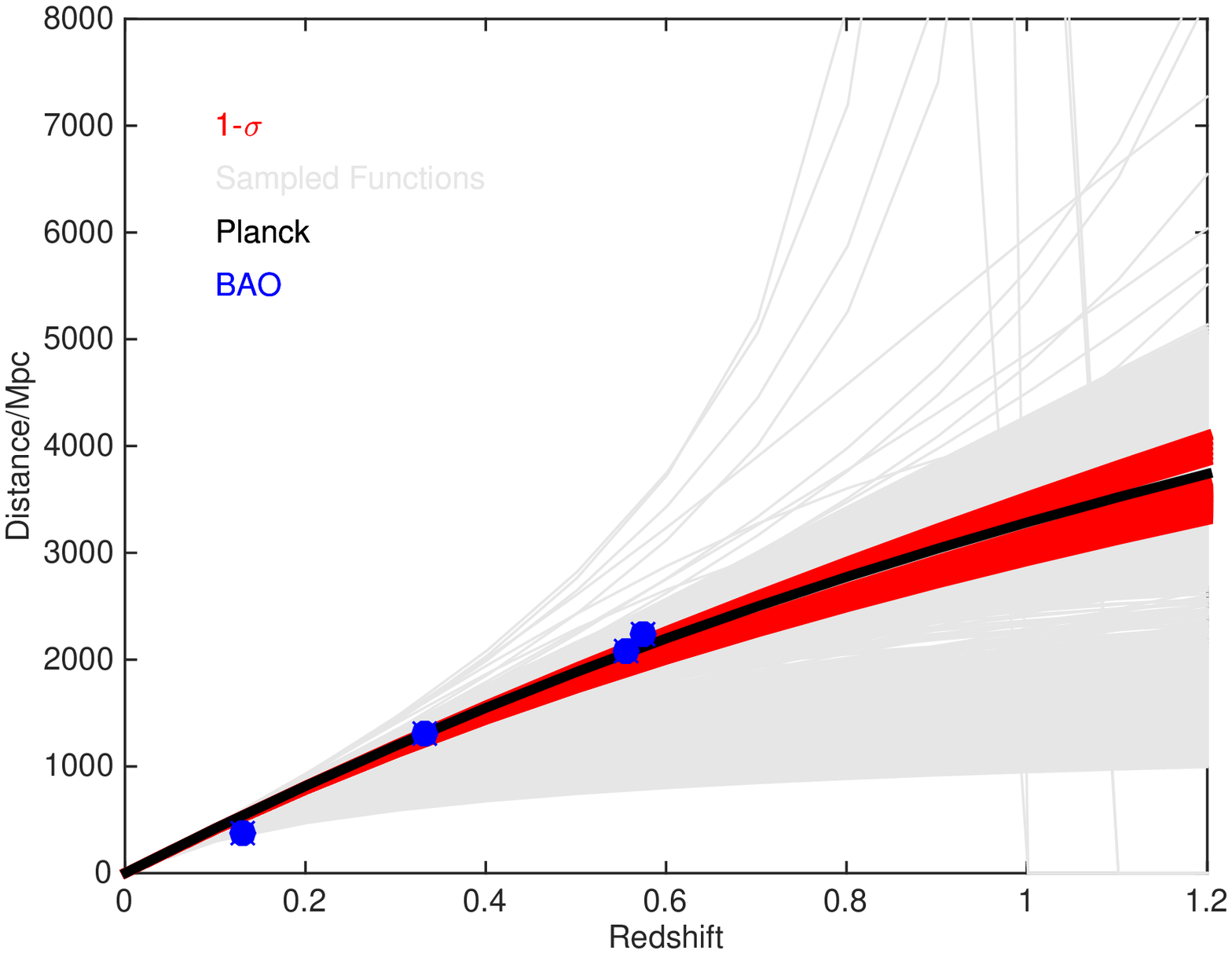}
  \includegraphics[angle=0,clip=,width=\columnwidth]{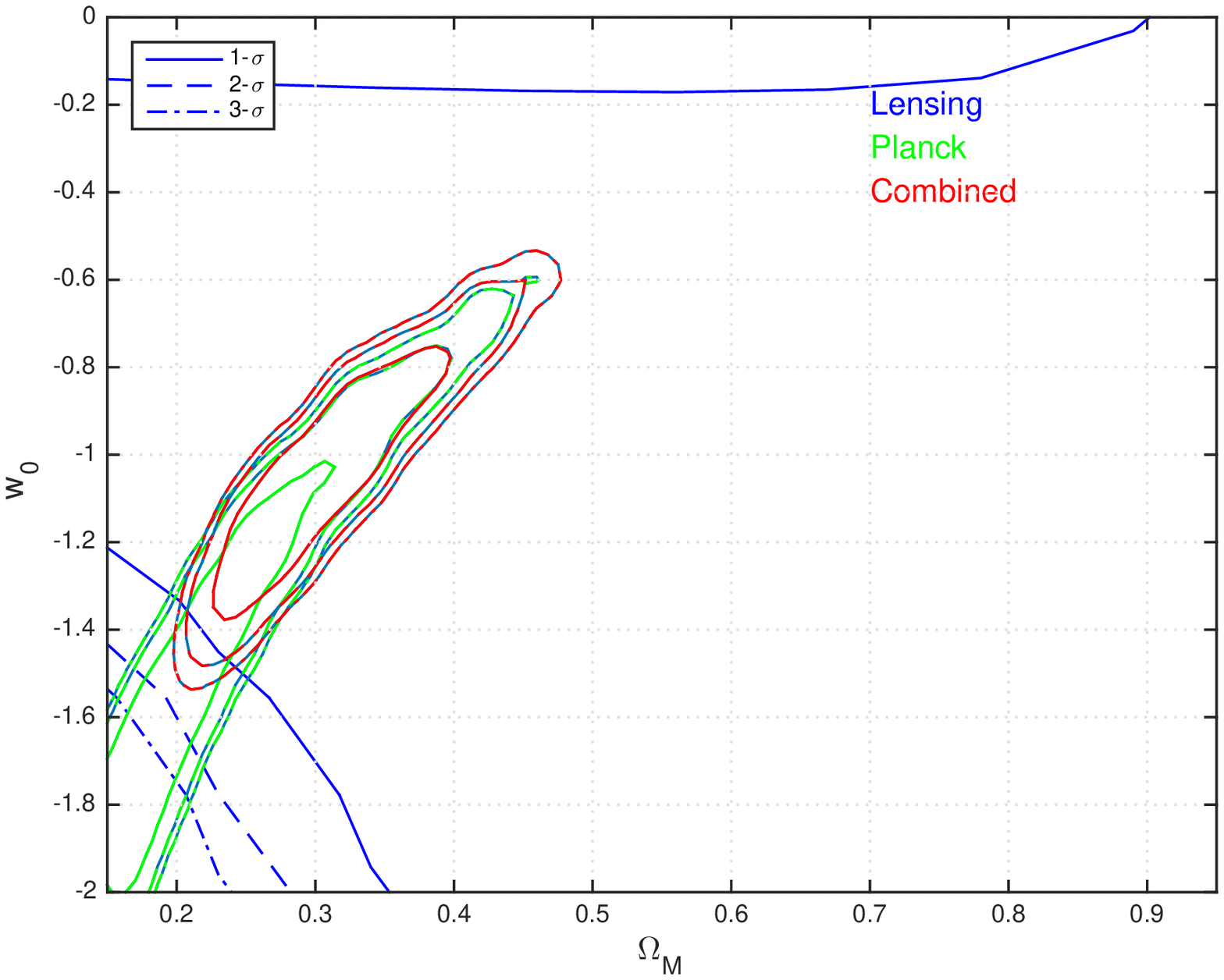}
 \caption{Left: The best fit comoving distance relation to the shear-ratio data. The grey 
lines show the set of functional forms considered, the solid red band shows the $1$-sigma 
confidence region about the maximum likelihood function. We use an anchor 
distance from NGC4258 (Humphreys et al., 2013). The black solid line shows the comoving distance inferred from the 
Planck best fit cosmology (Planck Collaboration, 2014; this is not a fit to the shear ratio data). 
The blue points show distances inferred from BAO data (6df, SDSS, BOSS and WiggleZ).
Right: Confidence limits on the dimensionless matter overdensity $\Omega_{\rm M}$ and a constant dark energy equation of state
$w_0$. The lines show the 2-parameter $1$, $2$, and $3$-sigma confidence limits for the shear-ratio method only (dark blue), Planck
only (green), and the combination (red).}
 \label{distance}
\end{figure*}

The cosmological parameter constraints from the shear-ratio method are not expected to be as small as 
from cosmic shear (see Taylor et al., 2007), but importantly it is only sensitive to the geometric information. Therefore, like 
other geometric-only methods, in combination with the CMB it should be able to lift degeneracies and improve constraints. 
Much like BAO and Type-1a supernovae constraints we can therefore also place a measurement on the distance-redshift relation. 
We test this by using a cosmological-parameter-free parameterisation of the comoving distance where 
$D(z)\propto z/\sqrt{az+bz^2}$ where $a$ and $b$ are free parameters (based on the fits in Pen, 1999\footnote{\tt https://www.ssl.berkeley.edu/$\sim$mlampton/ComovingDistance.pdf}). We fit these two parameters to the data using the likelihood 
in equation (\ref{like}), where we sample both $a$ and $b$ over the range $[0, 2]$ on a 2D $100\times 100$ grid, 
that we then supplement with $1000$ uniform random samples of $a$ and $b$ over the same range, which 
provides a good sampling of many functions including extreme solutions. We 
show the results in Figure \ref{distance}, where we show all sampled functions, and the $1-\sigma$ 
likelihood confidence region. The shear-ratio is not sensitive to 
the overall normalisation of the distance-redshift relation, therefore we need an `anchor' distance (similar to that required 
for Type-1a supernovae constraints). For the anchor redshift we use the 
the revised geometric maser distance to NGC4258 of Humphreys et al. (2013) which has a distance of  
$7.60\pm 0.22$ Mpc and a redshift of $0.0015$.

In Figure \ref{distance} we show the $1$-sigma confidence region about the best fit comoving distance-redshift relation. 
This is \emph{a fortiori} consistent with the relation inferred from the Planck best-fit cosmological 
model (Planck Collaboration, 2014), and measurements from BAO 
(6df, SDSS, BOSS and WiggleZ; Beutler et al 2011, Padmanabhan et al. 2012, Anderson et al. 2013, and Blake et al. 2011), 
and extends the geometry-only redshift relation to the $z\gs 1$ domain.  

We can also convert the ratio measurement into an inference on cosmological parameters 
(this is done directly on the data rather than through the non-parameteric distance relation). We 
assume a flat geometry, and the shear-ratio 
is not sensitive to the Hubble parameter normalisation $H_0$. Therefore, assuming a constant dark energy equation of state,  
the only free parameters in an Friedmann-Robertson-Walker cosmology are the 
dimensionless matter density $\Omega_{\rm M}$ and the 
dark energy equation of state parameter $w_0$. We use the Planck MCMC chains from the Planck 
Legacy Archive\footnote{Chain {\tt PLA/base\_w/planck\_lowl\_lowLike/base\_w\_planck\_lowl\_lowLike\_1}.}, 
and do not place any additional priors on any parameters. We note here that in the parameter estimation here we do not
assume the $H_0$ prior used above to derive the distance relation, where it is only required in that instance 
to use the shear ratio data to infer distances.

In Figure \ref{distance} we show the confidence levels for these parameters. 
Whilst the shear-ratio constraints themselves are relatively weak, there is a clear disfavouring at over $3$-sigma of the area 
around $\Omega_{\rm M}\simeq 0.1$ and $w_0\simeq -2.0$ (which causes a higher comoving distance of approximately $4000$ Mpc
at a redshift of $z\simeq 1.2$). The scaling of the shear ratio with cosmological parameters is 
described in Taylor et al. (2007). 
This is in the direction of the degeneracy of the Planck wCDM confidence region. Hence 
in combination with Planck the result is 
\ba 
\Omega_M=0.31 &\pm& 0.10\nn
w_0 = -1.02&\pm& 0.37, 
\ea
which is consistent with an $\Lambda$CDM model (i.e. $w_0=-1$). 
This is similar to the way in which CMB degeneracies are lifted by including 
BAO measurements (e.g. Beutler et al 2011, Padmanabhan et al. 2012, Anderson et al. 2013, Blake et al. 2011), 
and these results are consistent with the cosmological interpretation of these data. 
For example Aubourg et al. (2014) find $\Omega_{\rm M}=0.302\pm 0.008$ and $w_0=-0.97\pm 0.05$ with a combination of 
SDSS BAO, Union data set Supernovae with Planck data; and Parkinson et al. (2012) find $\Omega_{\rm M}=0.354\pm 0.041$ 
and $w_0=-1.215\pm 0.117$ using WiggleZ data in combination with Union supernovae and Planck. 

\section{Conclusion}
\label{Conclusion}
In this paper we present a measurement of the distance-redshift relation and inferred cosmological parameters using the 
shear-ratio method applied to the RCSLenS data. The method takes the ratio of the mean tangential shear signal about galaxy 
clusters at the same redshifts, and therefore any mass-dependency on the signal is cancelled-out, resulting in a statistic that 
is only sensitive to the geometric lensing kernel. 

Using a redshift anchor from NGC4258 (Humphreys et al., 2013) we 
measure the comoving distance using a flexible non-cosmological parametric model out 
to redshifts $z\gs 1$. In combination with Planck data we constrain the dimensionless matter overdensity and dark energy equation 
of state parameter to $\Omega_M=0.31 \pm 0.10$ and $w_0 = -1.02\pm 0.37$. 
We test our method by computing the cross-component shear signal from the same clusters and find this to be consistent with zero, 
and create some mock data from which we can recover the input cosmology. 
The main assumption that we make is that there is no redshift evolution of the mean lens mass within the lens redshift bins, 
and we tested that, given our statistical 
sensitivity, this does not impact our results. This supposition may not be true when using 
a larger sample of clusters, although thinner redshift slices could also be used in such a case, 
that would make the assumption more accurate. 

Extracting the geometric part of the weak lensing data allows for a probe that is complementary to cosmic shear 
studies that are sensitive to both the geometry of the Universe and the growth of structure. On small-scales of $\ls 1$Mpc 
cosmic shear studies are likely to be sensitive to baryonic feedback processes (Semboloni et al., 2011, Mead et al., 2015) and the non-linear growth of structure 
that are difficult to model. One way to mitigate such uncertainty is to remove those small-scales from the data (Simpson et al., 2013) 
or in the modelling (Kitching et al., 2014). This removal degrades cosmological parameter constraints, however 
the shear-ratio method on cluster and group scales offers a way to use the `cleaner' geometric part of the weak lensing signal 
to regain some of the cosmological information using weak lensing data alone. 
\\
%__________________________________________________________________

\noindent{\em Acknowledgements:} 
TDK is supported by a Royal Society University Research Fellowship. MV acknowledges support from the European Research Council under FP7 grant number 279396 and the Netherlands Organisation for Scientific Research (NWO) through grants 614.001.103.
HH is supported by an Emmy Noether grant (No. Hi 1495/2-1) of the Deutsche Forschungsgemeinschaft. 
DGG thanks the South African National Research Foundation for financial support. EvU acknowledges 
support from an STFC Ernest Rutherford Research Grant, grant reference ST/L00285X/1. AC and CH acknowledge 
support from the European Research Council under FP7 grant number 240185.
We are grateful to the wider RCS2 team for planning the survey, applying for observing time, and conducting the observations. We acknowledge use of the Canadian Astronomy Data Centre operated by the Dominion Astrophysical Observatory for the National Research Council of Canada's Herzberg Institute of Astrophysics. We would like to thank Matthias Bartelmann for being our external blinder, revealing which of the four catalogues analysed was the true unblinded catalogue at the end of this study. 
\newpage
%__________________________________________________________________

%_________________________________________________________________
\onecolumn
\section*{Appendix A: Lens Bin Width Tests} 
\begin{figure*}
\centering
  \includegraphics[angle=0,clip=,width=0.33\columnwidth]{ommw04.eps}
  \includegraphics[angle=0,clip=,width=0.33\columnwidth]{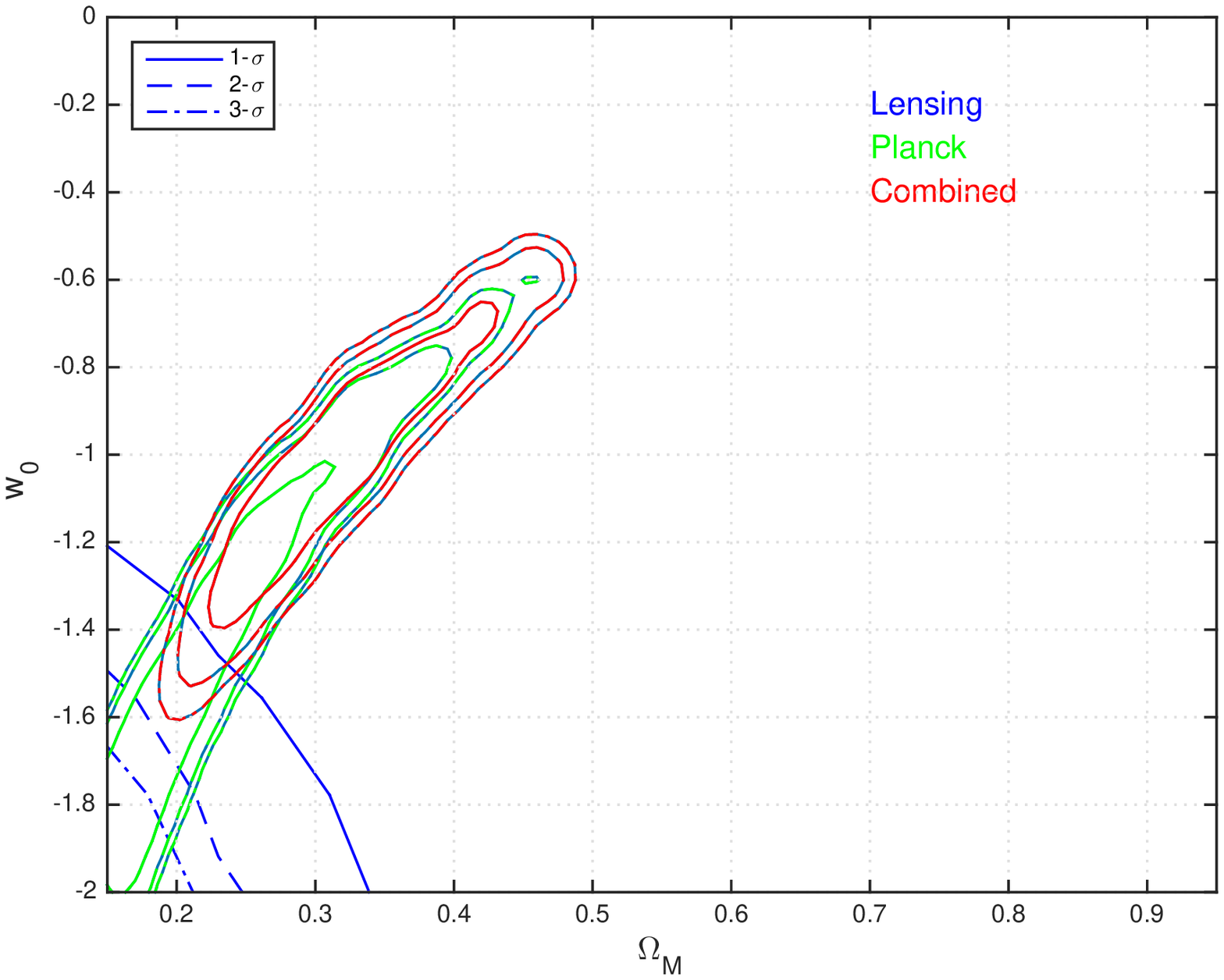}
  \includegraphics[angle=0,clip=,width=0.33\columnwidth]{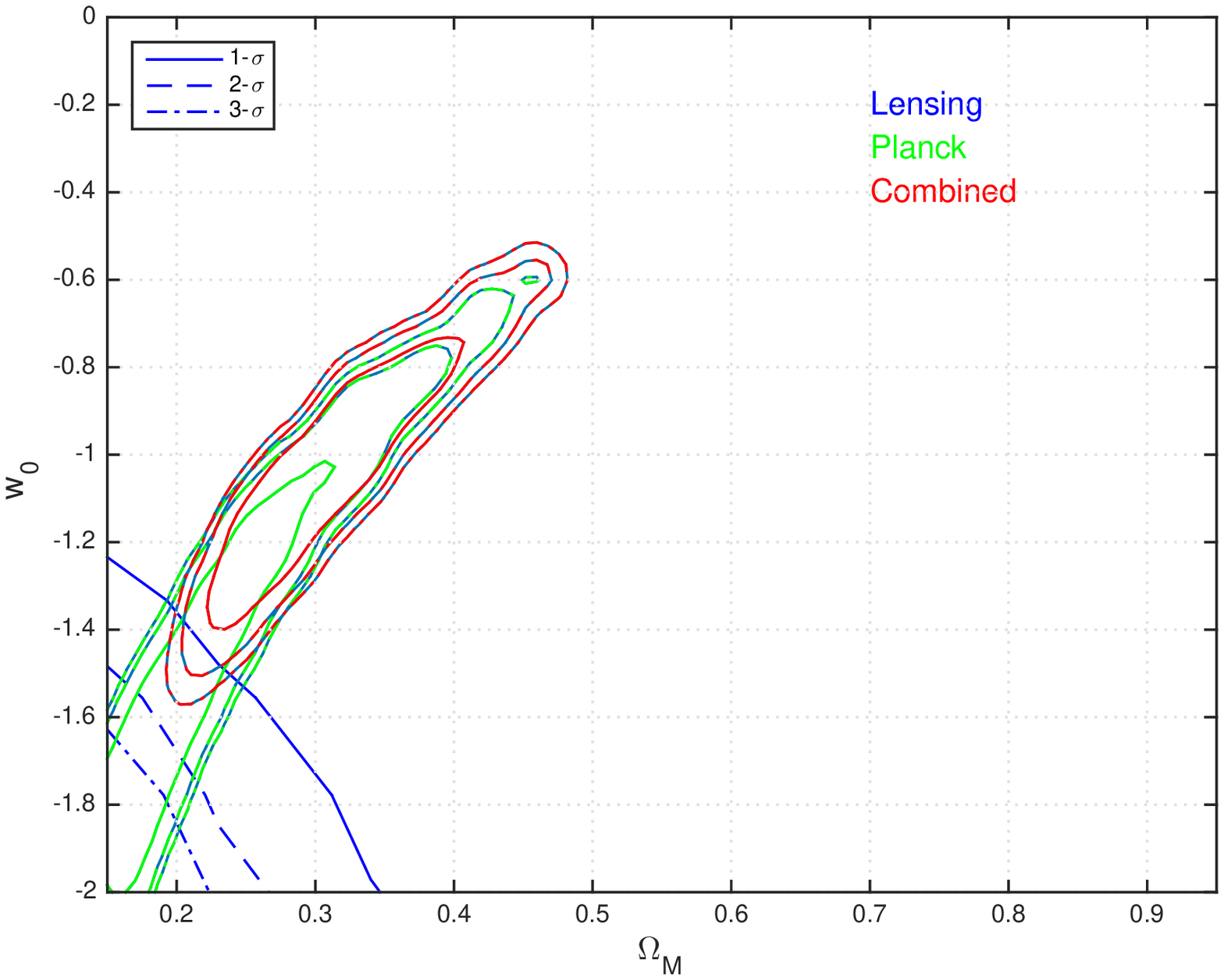}
 \caption{Cosmological constraints for three different lens bin widths. The left plot is the case used in the 
   main body of the paper $\Delta z_L=0.18$ corresponding to 2 bins over the range $0.3\leq z_L\leq 0.6$ (Figure 
   \ref{distance} on the righthand panel), 
   the middle plot is $\Delta z_L=0.09$ corresponding to 4 bins, and the right plot is for $\Delta z_L=0.045$ corresponding 
   to 8 bins.}
 \label{bins}
\end{figure*}

The width of the lens bins taken in this study affects various aspects on the analysis. The assumption that the mean 
mass dependency is constant as a function of redshift within a bin is correct for thin slices in redshift, but may affect the 
results for wider bins. In addition as the bin width is narrowed and more data points are used the results may be expected to 
improve as the resolution in redshift improves, but at the same time the error per data point increases, and the contamination 
between redshift bins due to the photometric redshift scatter also increases. In Figure \ref{bins} we show the cosmological 
parameter constraints obtained when varying the lens bin width. 
We find that all the results are statistically consistent. For our main conclusions we chose a conservative bin width of 
$\Delta z_L=0.18$ which has the fewest number of bins in the range we consider, and should be most robust to changes in 
the lens photometric redshift uncertainty. 

\section*{Appendix A: Simulation Tests}
\begin{figure*}
\centering
  \includegraphics[angle=0,clip=,width=0.49\columnwidth]{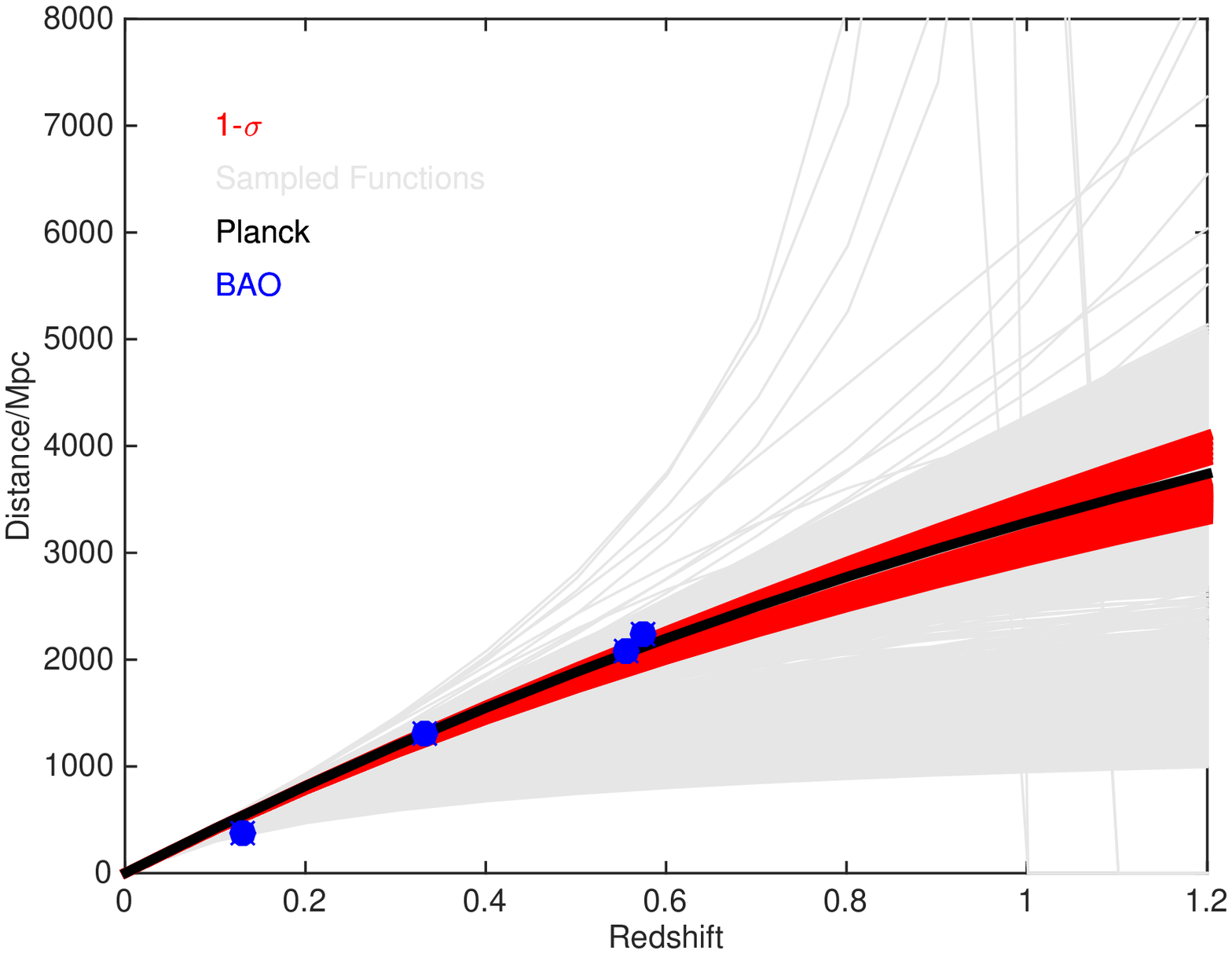}
  \includegraphics[angle=0,clip=,width=0.49\columnwidth]{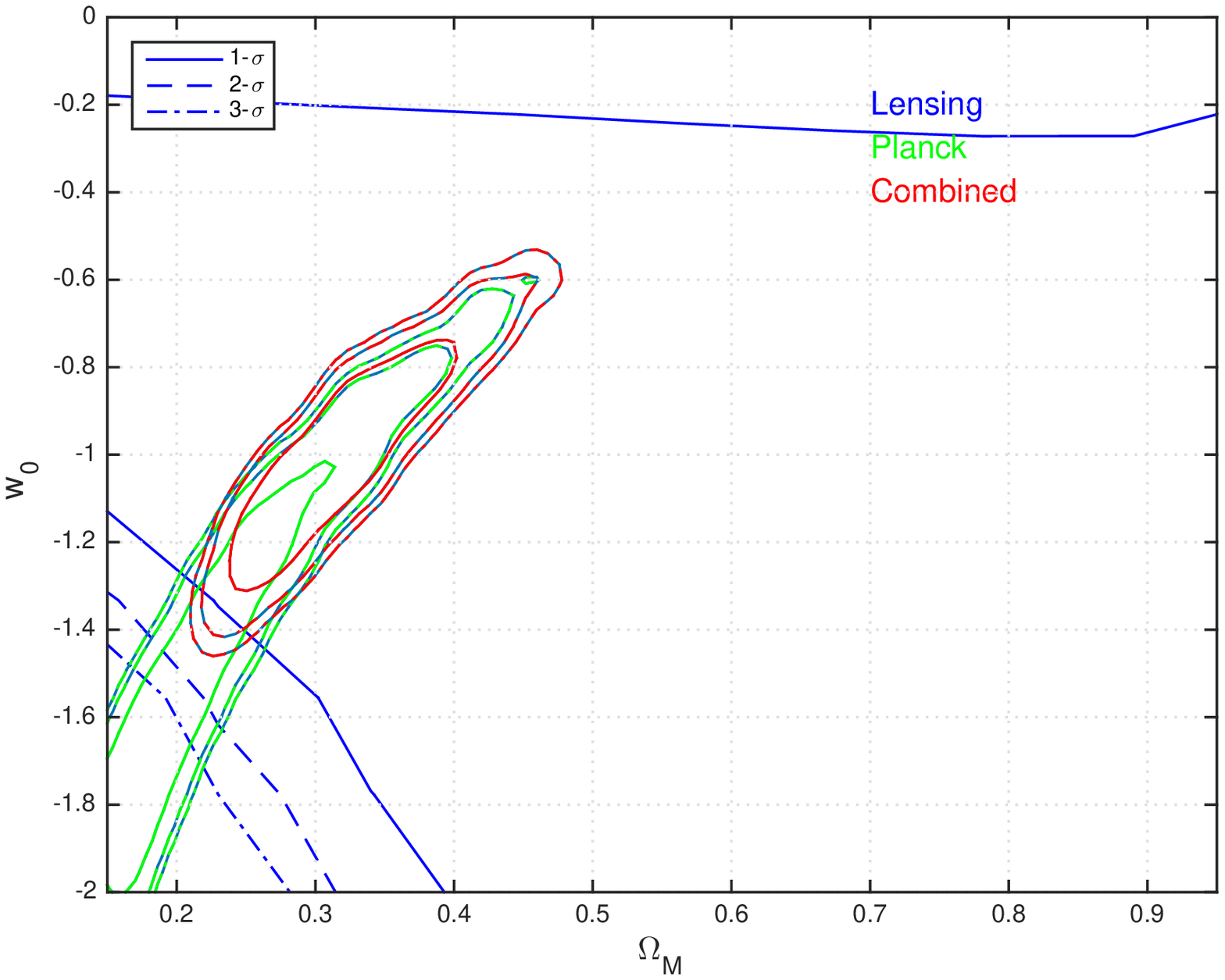}
 \caption{The best fit comoving distance relation to the mock shear-ratio data. The grey
lines show the set of functional forms considered, the solid red band shows the $1$-sigma
confidence region about the maximum likelihood function. We use an anchor
distance from NGC4258. The black solid line shows the comoving distance inferred from the
Planck best fit cosmology (Planck Collaboration, 2014). 
The blue points show distances inferred from BAO data (6df, SDSS, BOSS and WiggleZ). The right hand 
plot shows the inferred cosmological parameters, which are consistent with the input cosmology.}
 \label{sims}
\end{figure*}

To test the likelihood code, we create a set of ratios with a fiducial cosmology equal to the Planck best-fit cosmology, 
$\Omega_{\rm M}=0.32$ and $w_0=-1.0$, 
with a scatter corresponding to an error equal to the expected covariance; which is approximately $\sigma(R)\simeq 0.1$. 
This simulated the input into the likelihood and tests the assumption of the likelihood function. In Figure \ref{sims} we 
show that the input distance redshift relation and cosmology are recovered, and that the shape of the confidence regions 
is similar to that which we find with the data.

\end{document}